\documentclass[twocolumn]{aastex63}
\usepackage[nolist]{acronym}
\shorttitle{LIGO/Virgo S190828j and S190828l Are Not Strongly Lensed}
\shortauthors{Singer, Goldstein, \& Bloom}

\begin{document}

\title{The Two LIGO/Virgo Binary Black Hole Mergers on 2019 August 28 Were Not Strongly Lensed}

\author[0000-0001-9898-5597]{Leo P. Singer}
\affiliation{Astroparticle Physics Laboratory, NASA Goddard Space Flight Center, Code 661, 8800 Greenbelt Rd., Greenbelt, MD 20771, USA}
\email{leo.p.singer@nasa.gov}

\author[0000-0003-3461-8661]{Daniel A. Goldstein}
\altaffiliation{Hubble Fellow}
\affiliation{California Institute of Technology, 1200 E. California Blvd., Pasadena, CA 91125, USA}

\author[0000-0002-7777-216X]{Joshua S. Bloom}
\affiliation{Department of Astronomy, University of California, Berkeley, CA 94720-3411, USA}
\affiliation{Lawrence Berkeley National Laboratory, 1 Cyclotron Road, MS 50B-4206, Berkeley, CA 94720, USA}

\begin{abstract}
The LIGO/Virgo gravitational wave events S190828j and S190828l were detected only 21 minutes apart, from nearby regions of sky, and with the same source classifications (binary black hole mergers).
It is therefore natural to speculate that the two signals are actually strongly lensed images of the same merger.
However, an estimate of the separation of the (unknown) positions of the two events requires them to be $>10\arcdeg$ apart, much wider than the arcsecond-scale separations that usually arise in extragalactic lensing.
The large separation is much more consistent with two independent, unrelated events that occurred close in time by chance. We quantify the overlap between simulated pairs of lensed events, and use frequentist hypothesis testing to reject S190828j/l as a lensed pair at 99.8\% confidence.
\end{abstract}

\epsscale{1.15}

\section{Introduction}
\label{sec:intro}

On 2019 August 28 UT, the Advanced \acl{LIGO} (\acsu{LIGO}; \citealt{2015CQGra..32g4001L}) and the Virgo Gravitational-Wave Observatory (Virgo; \citealt{2015CQGra..32b4001A}) detected two \ac{GW} signals, S190828j and S190828l, that were separated in time by roughly 21 minutes \citep{GCN25497,GCN25503}.
Both signals were consistent with coming from \ac{BBH} mergers, and were both highly significant detections with false alarm rates of 1 per $3.7 \times 10^{13}$ years and 1 per $6.8 \times 10^2$ years, respectively.
Rapid localization using \texttt{BAYESTAR} \citep{2016PhRvD..93b4013S} revealed that the events had similarly-shaped localization probability contours, and that they appeared to originate from nearby regions of sky.
Together, these observations are suggestive of a possible strong lensing origin.

Recent articles have proposed that detections of strongly lensed \acp{GW} should be relatively common \citep{2018arXiv180205273B,2019arXiv190103190B}, while others have predicted that they should be relatively rare \citep{2017PhRvD..95d4011D, 2018PhRvD..97b3012N, 2018MNRAS.480.3842O, 2018MNRAS.475.3823S}.
Definitively establishing a strong-lensing origin for S190828j/l would be an important advance toward the resolution of this question.
By simulating \texttt{BAYESTAR} sky maps of lensed pairs of \ac{BBH} mergers and comparing them to the contents of the publicly available LIGO/Virgo alerts and localizations of S190828j/l, we show in this Letter that the two events are not the result of strong lensing, but rather are unrelated \ac{BBH} mergers that occurred close in time and relatively close in space.

\section{Strong Lensing}

In the strong-lensing interpretation of S190828j/l, the two events correspond to two lensed images of the same \ac{BBH} merger.
The images form due to the curvature of space-time by an unknown intervening mass along the line of sight.
The deflector may be a galaxy \citep[e.g.,][]{2006ApJ...638..703B}, a galaxy cluster \citep[e.g.,][]{2005ApJ...621...53B}, a dwarf galaxy \citep[e.g.,][]{2014Sci...344..396Q}, an extragalactic star \citep{1979Natur.282..561C}, an extragalactic field of stars \citep{1987A&A...171...49S}, or even a cosmic string \citep{1984MNRAS.211..575H}.
The image multiplicity is set by the mass distribution of the lens and its orientation with respect to the \ac{GW} source, and in principle can be much larger than two.
``Double'' or ``quad'' images are the most common products of galaxy-scale lensing \citep{2010ARA&A..48...87T}, while rich clusters routinely produce more than four images of background sources \citep[e.g.,][]{2016ApJ...821..116U}.
In the strong-lensing scenario, remaining images of S190828j/l, if they exist, were either too faint to trigger alerts, arrived when the interferometers were not in observing mode, or may still be yet to arrive.

In the strong-lensing scenario, the lensed images S190828j and S190828l travel along different geometric paths and through geometric potentials to reach us, explaining their 21 minute difference in arrival times.
This time-delay effect is a key feature of strong lensing \citep{1964MNRAS.128..295R}.
Observed time delays in strong lensing systems have ranged from minutes to hours at the low end \citep[e.g., SN iPTF16geu;][]{2017Sci...356..291G,2017ApJ...835L..25M,2019arXiv190706609M}, to decades at the high end \citep[e.g., SN Refsdal;][]{2015Sci...347.1123K}.
A strong-lensing hypothesis thus seems, at broad brush, to offer a convenient explanation for the spatial and temporal proximity of S190828j and S190828l.
However, as we will show in this Letter, the spatial separation between S190828j and S190828l ($>$10$^\circ)$ turns out to be too large and the time delay (21 minutes) too small for lensing to work.

\subsection{Effect of Lensing on LIGO Observables}

Before presenting our analysis of the LIGO data, we first review the effects of gravitational lensing on \ac{GW} strain signals and present some useful scaling relations.
In an expanding universe, the leading-order post-Newtonian approximation to the \ac{GW} strain from a compact binary merger is (cf. \citealt{1986Natur.323..310S,2005ApJ...629...15H,2010ApJ...725..496N})
\begin{equation}
    h(f) = \Theta \frac{(1 + z) \mathcal{M}}{D_\mathrm{L}} (\pi (1 + z) \mathcal{M} f)^{2/3} e^{-i\Psi(f)}.
\end{equation}
Here, the factor $\Theta$ encapsulates all of the dependence on the orientation and sky position of the binary, $z$ is the redshift of the binary, $f$ is the \ac{GW} frequency, $D_\mathrm{L}$ is the luminosity distance, $\Psi(f)$ is the leading-order phase as a function of frequency, and $\mathcal{M}$ is the \emph{chirp mass} defined in terms of the component masses as $\mathcal{M} = (m_1 m_2)^{3/5} (m_1 + m_2)^{-1/5}$.

One can recast this in terms of the \emph{redshifted chirp mass} $\mathcal{M}_z = (1 + z) \mathcal{M}$ (and similarly the redshifted component masses $m_{1,z} = (1 + z) m_1$ and $m_{2,z} = (1 + z) m_2$) as
\begin{equation}
    h(f) = \Theta \frac{\mathcal{M}_z}{D_\mathrm{L}} (\pi \mathcal{M}_z f)^{2/3} e^{-i\Psi(f)}.
\end{equation}
The redshifted masses are also referred to as the \emph{observer frame} masses, in contrast to the physical \emph{source frame} masses.
Strong lensing does not alter the redshifted, observer-frame masses.
The apparent luminosity distance, however, is modified by the lensing magnification $\mu$ according to
\begin{equation}
    D_L^\prime = \frac{D_L}{\sqrt{|\mu|}}.
\end{equation}

\subsection{Angular and Temporal Scales for Lensing}

The angular scale of separation between a pair of lensed images is given by the Einstein angle $\theta_\mathrm{E}$, which in the case of a point lens takes the form
\begin{equation}
\label{eq:einsteinrad}
    \theta_\mathrm{E} = \left(\frac{4GM}{c^2}\frac{D_{ls}}{D_lD_s}\right)^{1/2},
\end{equation}
where $M$ is the mass of the lens, $D_l$ is the angular diameter distance to the lens, $D_s$ is the angular diameter distance to the source, and $D_{ls}$ is the angular diameter distance between the lens and the source \citep{1992grle.book.....S}.
In addition to bound structures, cosmic strings, hypothesized one-dimensional topological defects in spacetime, may also produce lensing, but the image separation has a different dependence on distance than in Equation~\ref{eq:einsteinrad} \citep{2003PhRvD..68b3506P,2010MNRAS.406.2452M}.
The Einstein angle $\theta_E$ of a cosmic string is given by
\begin{equation}
    \label{eq:strings}
    \theta_\mathrm{E} = \frac{8 \pi G \mu}{c^2} \sin i \frac{D_{ls}}{D_s},
\end{equation}
where $\mu$ is the string ``tension'' and $i$ is the angle that the cosmic string forms with the line of sight.
There are observational limits on cosmic string tension $G \mu / c^2$ from optical surveys \citep{2010MNRAS.406.2452M}, the cosmic microwave background \citep{2014A&A...571A..25P}, and even \acp{GW} \citep{2018PhRvD..97j2002A}.

\begin{figure}
    \includegraphics[width=\columnwidth]{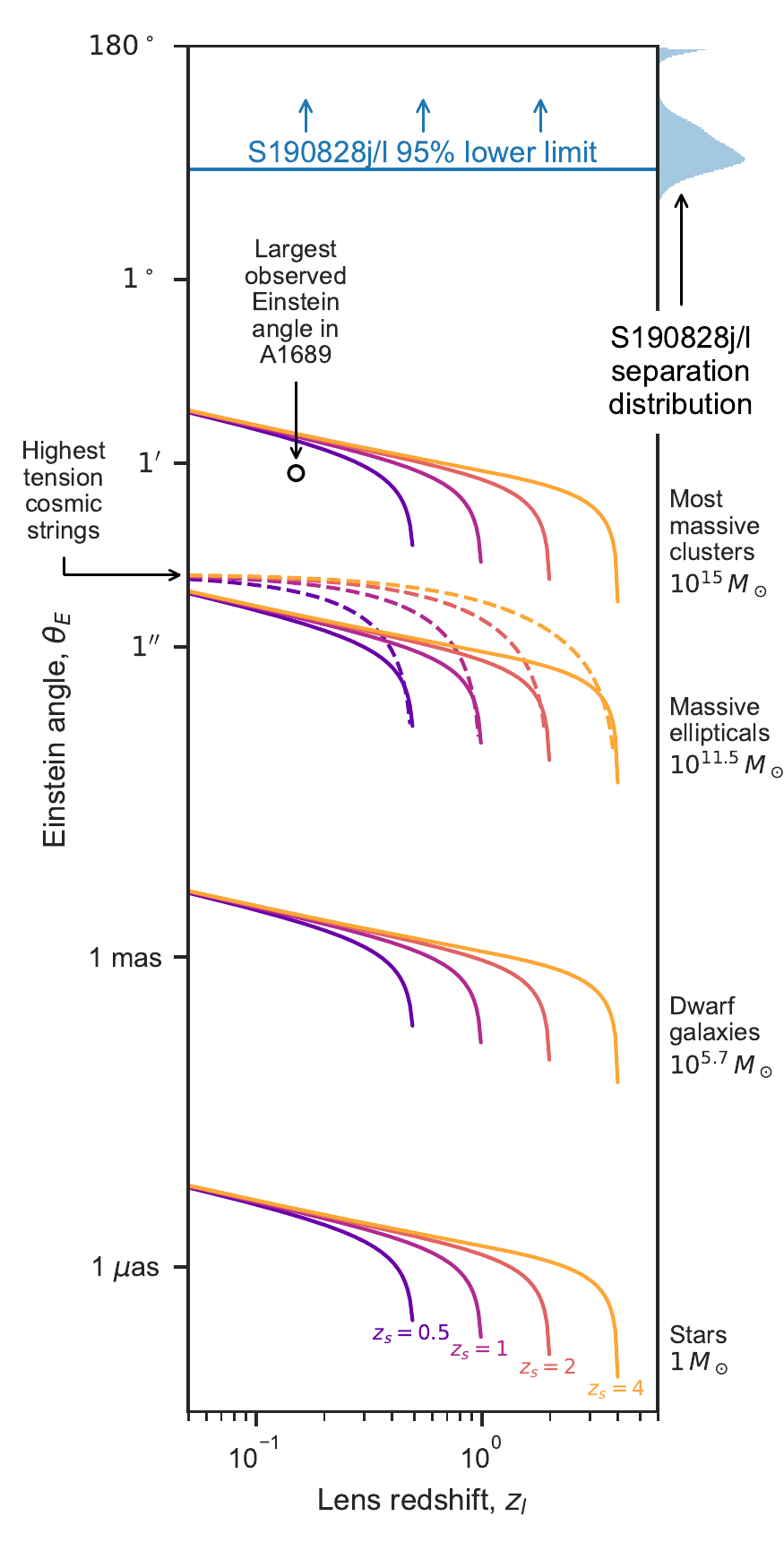}
    \caption{\label{fig:lensing}Einstein angle $\theta_\mathrm{E}$ as a function of lens redshift $z_l$ of a variety of strong lensing systems: extragalactic stars or star fields (``Stars''), dwarf galaxies, elliptical galaxies, and superclusters, the most massive bound structures in the universe.
The posterior probability distribution of the separation of S190828j and S190828l is shown on the right (also in the top panel of Figure~\ref{fig:separation-magnification}), and the 95\% lower limit on the separation is shown as the horizontal blue line near the top.}
\end{figure}

Dimensionally, Equation~\ref{eq:einsteinrad} can be expressed as
\begin{equation}
\label{eq:dimensional}
    \theta_\mathrm{E} = \left(\frac{M}{10^{11.09}M_\odot}\right)^{1/2}\left(\frac{D_lD_s/D_{ls}}{\mathrm{Gpc}}\right)^{-1/2}\,\mathrm{arcsec}.
\end{equation}
Similarly, setting $\sin i = 1$ in Equation \ref{eq:strings} for the greatest possible magnification, we obtain
\begin{equation}
    \label{eq:stringdimensional}
    \theta_\mathrm{E} = 5.2 \left(\frac{G \mu / c^2}{10^{-6}}\right) \left(\frac{D_{ls}}{D_s}\right)\,\mathrm{arcsec}.
\end{equation}
Plugging numbers into Equations \ref{eq:dimensional} and \ref{eq:stringdimensional} gives the spatial scale for lensing in various mass and distance regimes, which we have plotted in Figure~\ref{fig:lensing}, assuming a \cite{2016A&A...594A..13P} cosmology.
By comparing the angular separation S190828j/l sky maps to the scales for lensing in Figure \ref{fig:lensing} across 15 decades in mass, we will show in the next section that the lensing separations associated with even the most massive bound structures in the universe are many orders of magnitude smaller than those required to explain S190828j/l.

Finally, in most plausible astrophysical lensing scenarios, such as galaxy-galaxy lensing, galaxy-cluster lensing, and extragalactic stellar microlensing, time delays increase as image separations increase.
This can be understood as a consequence of geometry: for fixed source and lens positions, larger image separations (due e.g., to increasing the lens mass) lead to larger differences in path length, which in turn lead to larger time delays.
For a given lens, the magnitude of the time delay between two lensed images depends sensitively on the location of the unlensed source relative to the lens.
For microlensing by stars, typical time delays are a few microseconds \citep{1996IAUS..173..279M}; for galaxy-galaxy lensing, time delays typically range from a few hours to a few months \citep{2010MNRAS.405.2579O,2017ApJ...834L...5G}; and for cluster lensing, time delays can be as high as several decades \citep[e.g.,][]{2015Sci...347.1123K}.
As Figure \ref{fig:lensing} shows, the $\gtrsim 10^\circ$ separation of S190828j/l requires a lens with a mass roughly $10^6$ times that of the largest known bound structures in the Universe.
As larger image separations in general correlate with larger time delays, achieving a 21 minute time delay with such a large lens would require exceptional source-lens alignment, assuming a spherically symmetric mass distribution.
The short time delay and large separation of S190828j/l futher strain the lensing interpretation.

\section{Analysis of the public LIGO/Virgo data}
\label{sec:results}

\begin{figure}
\includegraphics[width=\columnwidth]{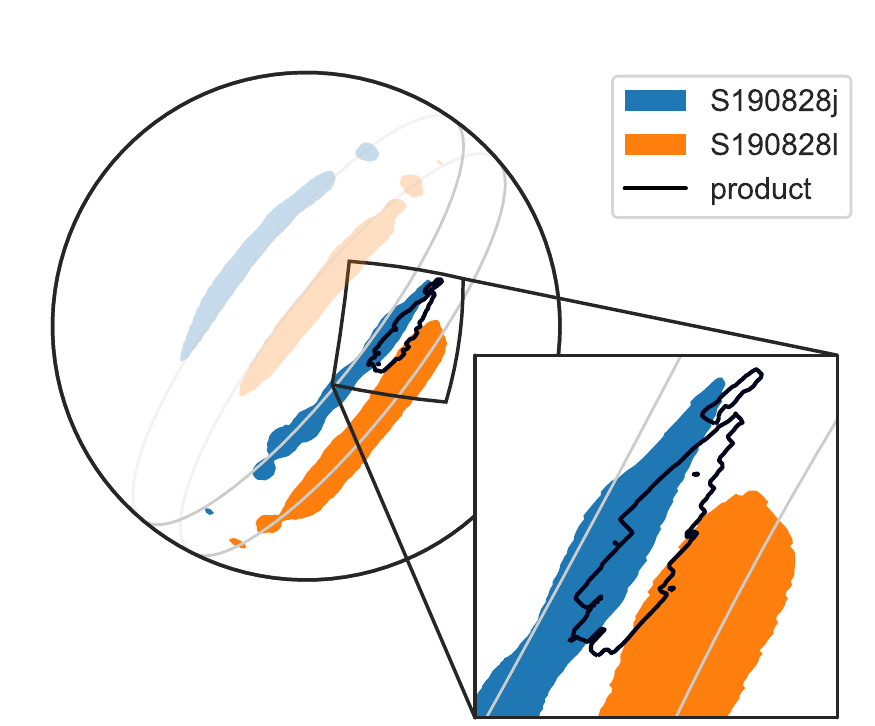}
\caption{\label{fig:skymaps}The \texttt{BAYESTAR} 90\% credible localization regions of the two mergers and their joint product.
The two parallel gray lines are lines of constant time delay between Hanford and Livingston and are separated by 13$\arcdeg$.
The 90\% credible areas of S190828j and S190828l are 587 and 948\,deg$^2$ respectively, and the 90\% credible region of their joint product is 94\,deg$^2$.
For \texttt{LALInference}, the events' 90\% credible areas are 228 and 359\,deg$^2$ respectively, with a joint 90\% credible area of 11\,deg$^2$.}
\end{figure}

To test the strong lensing hypothesis described in the previous section, we downloaded the initial \texttt{BAYESTAR} localizations \citep{2016PhRvD..93b4013S,GCN25497,GCN25503}  of S190828j and S190828l and plotted them in Figure~\ref{fig:skymaps}. We also inspected the refined \texttt{LALInference} localizations \citep{2015PhRvD..91d2003V,GCN25782,GCN25861}.

Each of the localizations is bimodal, consisting essentially of an annulus determined by the time delay on arrival at Hanford and Livingston, carved into two opposing segments by the sensitivity nulls in the Hanford and Livingston antenna patterns.
The 90\%-credible regions of the two localizations do not overlap.

Nonetheless, if we assume that they are lensed images with negligible separation compared to the LIGO/Virgo localization uncertainty, then we can form a joint localization region by multiplying and then normalizing the two sky maps.
Denoting the two localizations as $p(\hat{\mathbf{n}})$ and $q(\hat{\mathbf{n}})$, the joint localization is
\begin{equation}
r(\hat{\mathbf{n}}) = \frac{p(\hat{\mathbf{n}}) q(\hat{\mathbf{n}})}{\int p(\hat{\mathbf{n}}^\prime) q(\hat{\mathbf{n}}^\prime) \, d\Omega^\prime}.
\end{equation}
Operationally, since the sky maps are stored discretely as normalized arrays of equal-area HEALPix%
\footnote{\url{https://healpix.sourceforge.io}}
pixels, this is evaluated as
\begin{equation}
r_i = \frac{p_i q_i}{\sum_j p_j q_j},
\end{equation}
where the sum is over pixels.
This joint localization is shown as the black contour in Figure~\ref{fig:skymaps}.
The denominator of this expression can be interpreted as a Bayes factor comparing the lensed and unlensed hypotheses \citep{2018arXiv180707062H,2019ApJ...874L...2H}, and is equal to
\begin{equation}
\mathcal{B}^L_U = \int p(\hat{\mathbf{n}}^\prime) q(\hat{\mathbf{n}}^\prime) \, d\Omega^\prime = {\sum_i p_i q_i} = 3.9 \times 10^{-9},
\end{equation}
disfavoring lensing at about the $6$-$\sigma$ level.
For \texttt{LALInference}, the Bayes factor of $\mathcal{B}^L_U = 5.7 \times 10^{-10}$ rejects lensing even more strongly.

\begin{figure}
\includegraphics[width=\columnwidth]{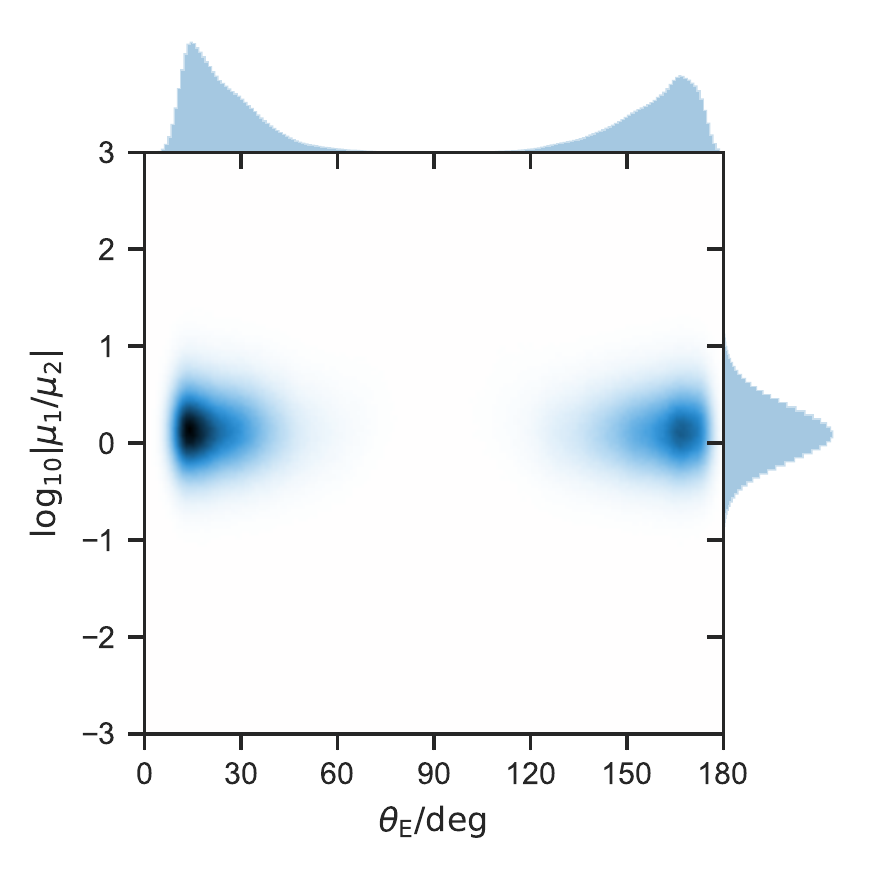}
\caption{\label{fig:separation-magnification}Joint posterior probability distribution of the relative magnification and separation of the images.
The separation is constrained to be $>1$ deg at $\approx 99.99\%$ confidence.
The relative magnification is consistent with unity, with an uncertainty of about half a decade in either direction.}
\end{figure}

Alternatively, we can assume that the events are lensed images but with no prior assumption about the separation.
We can then calculate the joint probability distribution of the image separation $\theta$ and relative magnification $\left|\mu_1 / \mu_2\right|$.
In the geometric optics limit that produces multiple distinct images, lensing is achromatic and therefore it does not alter the apparent detector-frame masses.
Its only impacts are to alter the arrival time, sky location, and apparent luminosity distance of the signal.
Therefore $\left|\mu_1 / \mu_2\right| = (D_{\mathrm{L},2} / D_{\mathrm{L},1})^2$.
Assuming a uniform prior on sky location, the posterior distribution of separation and relative magnification is shown in Figure~\ref{fig:separation-magnification}.
The relative magnification is consistent with unity; it is roughly log-normally distributed such that $\log_{10} \left|\mu_1 / \mu_2\right| = 0.10^{+0.56}_{-0.53}$.%
\footnote{All quantities in this paper written in the form $x_{-y}^{+z}$ have a median value of $x$ and a 5\% to 95\% credible interval of $[x-y, x+z]$.}
For \texttt{LALInference}, the relative magnification is $\log_{10} \left|\mu_1 / \mu_2\right| = 0.21_{-0.50}^{+0.54}$.

The separation distribution has two modes containing comparable probability mass, corresponding to the possibilities that the two images are on either adjacent or opposite sections of the triangulation rings. The adjacent mode is favored over the opposite mode by a ratio of $1.2:1$.
The first mode is at $\theta = 22\arcdeg^{+27\arcdeg}_{-11\arcdeg}$ and the second is at $\theta = 161\arcdeg^{+13\arcdeg}_{-31\arcdeg}$.
There is only a $\approx 0.01\%$ chance that the separation is smaller than 1$\arcdeg$. For the \texttt{LALInference} localization, the adjacent mode is favored over the opposite mode by a factor of $6.3:1$. The two modes are at $\theta = 19\arcdeg_{-10\arcdeg}^{+20\arcdeg}$ and $\theta = 160\arcdeg_{-38\arcdeg}^{14\arcdeg}$, respectively, and there is only a $\approx 0.007\%$ chance that the separation is smaller than $1\arcdeg$.

\section{Analysis of Simulated Localizations}
\label{sec:sims}

To evaluate whether S190828j and S190828l are consistent with a strong lensing origin, we carried out a suite of simulations providing an estimate of the distribution of sky maps that LIGO/Virgo would have observed if:

\begin{enumerate}
    \item H1, L1, and V1 detected two \ac{GW} events at the exact times of S190828j and S190828l, corresponding to two lensed images of the same \ac{BBH} merger.
    \item H1, L1, and V1 detected two \ac{GW} events at the exact times of S190828j and S190828l, corresponding to independent \ac{GW} signals produced by two unrelated \ac{BBH} mergers.
\end{enumerate}

The key difference between Cases 1 and 2 is that in Case 1, the true sky locations and the signal-to-noise ratios of the \ac{GW} signals are tightly correlated due to lensing, whereas in Case 2, they are independent.

\subsection{Detector Sensitivity}

The sensitivities of the \ac{GW} detectors in our simulation were matched to the performance of LIGO and Virgo around the time of S190828j/l.
LIGO/Virgo do not publish live noise curves.
However, the Gravitational Wave Open Science Center's Gravitational-Wave Observatory Status page for August 28%
\footnote{\url{https://www.gw-openscience.org/detector_status/day/20190828/}}
gave the binary neutron star range between the times of S190828j and S190828l as follows: H1, 113\,Mpc; L1, 135\,Mpc; V1, 47\,Mpc.
To approximate the sensitivity of the LIGO and Virgo detectors around the time of the two events, we took the \texttt{aLIGOMidHighSensitivityP1200087} and \texttt{AdVMidHighSensitivityP1200087} noise curves from \texttt{LALSimulation} and applied constant scale factors to them in order to match the aforementioned binary neutron star range.

\subsection{Source Distribution}

We obtained the distribution of source parameters by simulating a uniform-in-comoving-volume population of binary black hole mergers with source frame component masses distributed uniformly in log mass in $[5, 50]$\,$M_\odot$ and aligned component spins uniformly distributed in $[-0.99, 0.99]$.
This matches the ``uniform in log'' distribution used to estimate \ac{BBH} merger rates in \citet{PhysRevX.9.031040}.
It does not matter that the source population assumes a particular mass function and no lensing, because it is simply a construct to generate a set of binaries with \emph{observer frame} masses that are broadly consistent with what LIGO/Virgo has observed.
Source locations are isotropic but confined to the intersection of the 99.9999999997\% credible regions of S190828j and S190828l (with an area of 6800\,deg$^2$) to be broadly consistent with the observed positions of the two events.
Signals were injected into Gaussian noise and recovered using matched filtering.
Only events that registered an SNR $>4$ in at least 2 detectors and a network SNR $>12$ were kept. For every surviving event, we ran sky localization with \texttt{BAYESTAR}. (We did not run \texttt{LALInference} because it would have been computationally prohibitive.)

\subsection{Strongly Lensed Events Simulation}

The lensed population is constructed as follows.
We draw $10^5$ independent samples from the source parameter distribution.
Each sample is injected into a stretch of Gaussian noise at a fixed time $t_0$.
Then the sample is moved to a new random sky location that is drawn uniformly from a cone with a radius of $30\arcsec$ centered on the old sky location, consistent with the largest observed lensing separations from Figure~\ref{fig:lensing}.
The new apparent luminosity distance is drawn from a log-normal distribution with a width of 0.25\,dex centered on the old distance, consistent with the 0.5\,dex scatter in relative magnification inferred in Figure~\ref{fig:separation-magnification}.
It is then injected again into another independent stretch of Gaussian noise at a fixed time $t_0 + 21\,\mathrm{min}$.
Pairs of events that pass the detection thresholds at both sidereal times are kept.
The lensed population consists of $N_\mathrm{lensed} = 696$ pairs of events.

\subsection{Independent Events Simulation}

The independent events population is constructed in a similar manner.
We draw $10^5$ independent samples form the source parameter distribution and inject them at a time $t_0$, and keep the ones that pass the detection thresholds.
Then we draw another $10^5$ independent samples and inject them at a time $t_0 + 21\,\mathrm{min}$.
Events from the first and second group of $10^5$ that pass the detection thresholds are matched pairwise.
The independent events population consists of $N_\mathrm{independent} = 989$ pairs of events.

\subsection{Frequentist Hypothesis Testing}

We employ two closely related test statistics to quantify the overlap between each pair of events.
The first is the Bhattacharyya coefficient \citep{Bhattacharyya}, $F(p, q) = \sum_i \sqrt{p_i q_i}$, which in the application to \ac{GW} localizations has also been referred to as the \emph{fidelity} \citep{2017MNRAS.466L..78V}.
The second is the Bayes factor defined in Section~\ref{sec:results}, defined as $G(p, q) = \sum_i p_i q_i$.

\begin{figure*}
\plottwo{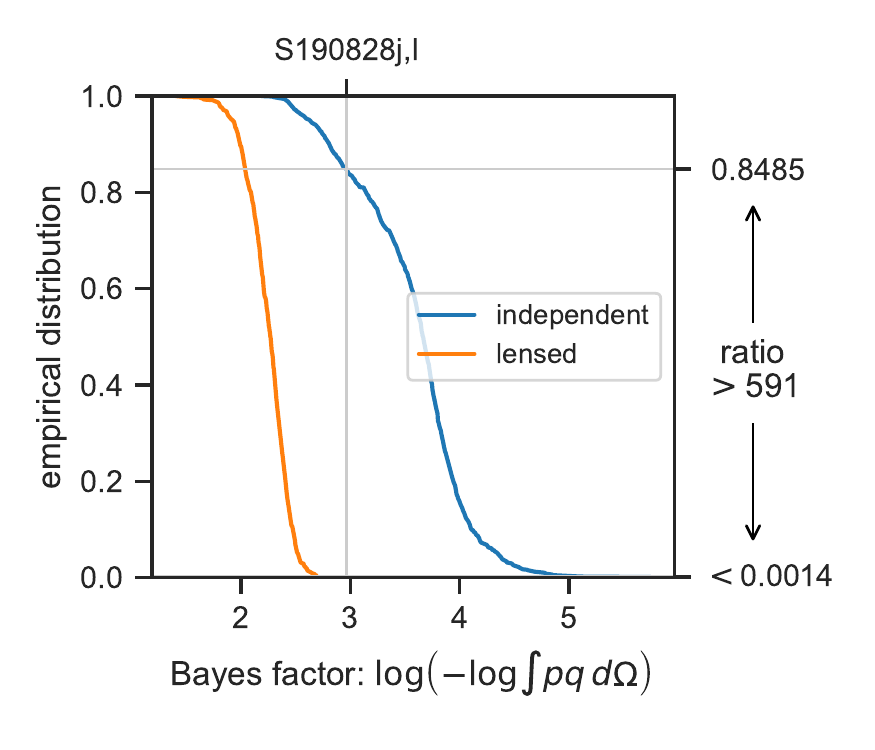}{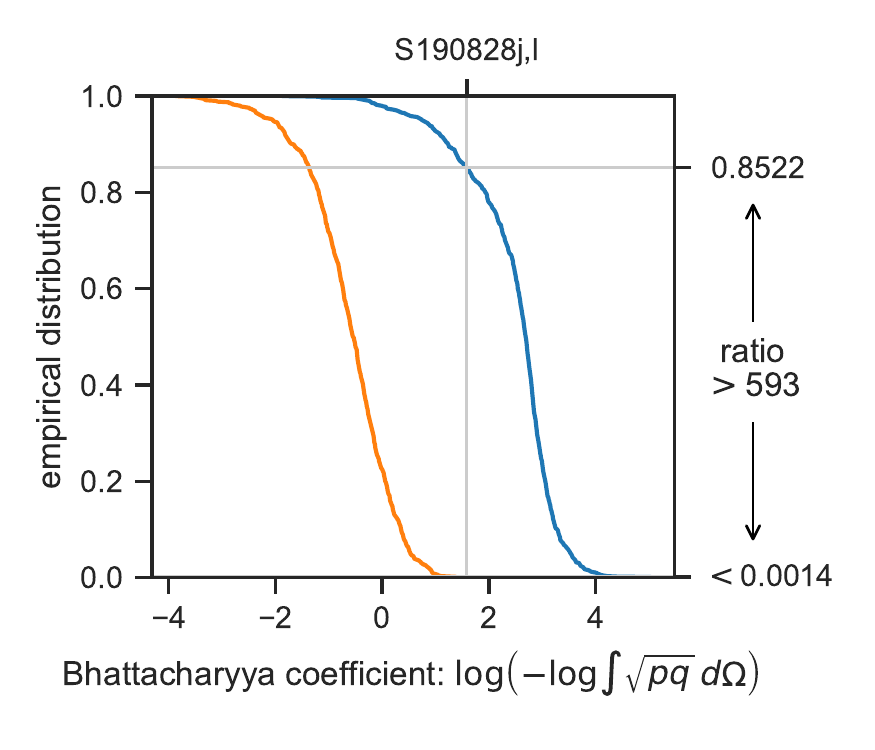}
\caption{\label{fig:frequentist}Empirical distribution of the test statistics (left panel: Bayes factor; right panel: Bhattacharyya coefficient) for a population of simulated lensed binary black hole mergers and for a pairs of independent events.
The horizontal scale is doubly logarithmic.
The values of the test statistics for S190828j/l are shown by the vertical gray lines.
$P$-values for S190828j/l under the lensed and independent events hypotheses are shown as horizontal gray lines.}
\end{figure*}

We tabulated both test statistics for each pair of simulated events.
Figure~\ref{fig:frequentist} shows the empirical distribution function of the test statistics $F(p, q)$ and $G(p, q)$ over the simulated events.
The value of the test statistics for the S190828j/l pair is shown as a vertical gray line.
The corresponding $P$-values for the lensed and independent events hypotheses are shown as horizontal gray lines.

The $P$-values for both test statistics under the independent events hypothesis happen to be the same to two significant digits, about $P_\mathrm{independent} = 0.85$, very much consistent with S19082j/l being independent events.

None of our lensed simulations had values of the test statistics that were as extreme as S190828j/l.
Therefore we can only provide an upper bound on the P-value for the lensed hypothesis of $P_\mathrm{lensed} < 1 / N_\mathrm{lensed} = 0.0013$, inconsistent with the lensed hypothesis.
Performing a larger number of simulations would only strengthen the rejection of the lensed hypothesis.

\section{Interpretation}
\label{sec:interpretation}

It may have been tempting to interpret S190828j and S190828l as strongly lensed images of the same event due to their similar times, similarly shaped localizations, and similar directions on the sky.
We offer the following responses to those points.

\paragraph{1. The time coincidence, while striking, is not highly significant by itself.}
From April through September, LIGO/Virgo released 21 \ac{BBH} merger alerts, at an average rate of about $R=0.75$\,week$^{-1}$.
Let us assume that the rate of \ac{BBH} detections is time-independent.
In reality the detection rate does fluctuate with the sensitivity and uptime of the detector network, but accounting for the time variation will not have a significant impact on this line of reasoning.
The Poisson probability of $k=2$ or more detections occurring by coincidence during the \emph{same period} of $T = 21$\,min is low, $P(k \geq 2) = 1 - e^{-RT} (1 + RT) = 1.2 \times 10^{-6}$.
However, the \emph{rate of pairs of events} separated by time $T$ is roughly $2 P(k \geq 2) / T \approx .061$\,year$^{-1}$, or one in 16 years.
(The rate of double hits was verified by Monte Carlo simulation).
After several years of operation, it requires only a little bit of luck for Advanced LIGO/Virgo to observe two events separated by 21 minutes.

\paragraph{2. The localizations look qualitatively similar because the events occurred at similar sidereal times.} The qualitative features of \ac{GW} localizations are strongly determined by the positions and orientations of the detectors, which rotate with the earth.
This is especially true of the present \ac{GW} detector network, which consists of the two LIGO detectors with nearly aligned antenna patterns, and the much less sensitive Virgo detector.
The vast majority of localizations fit the description of a long, thin arc, almost a great circle, arising from the time delay on arrival between Hanford and Livingston, with two polar opposite sections above the antenna pattern maxima favored, and two polar opposite sections that are above the antenna pattern minima strongly disfavored \citep{2014ApJ...795..105S}.
One can see this pattern in Advanced LIGO and Virgo's first and second observing run at a glance in Figure~3 of \citet{2019ApJ...875..161A} or Figure~8 of \citet{PhysRevX.9.031040}, and even more clearly in Figure~2.5 of \citet{2015PhDT.........6S} by plotting the localizations of many events in Earth-fixed coordinates.
S190828j and S190828l definitely fit the mold.
Their localizations look so similar in celestial coordinates \emph{because they occurred at similar sidereal times}.
Their localizations would be likely to look this similar \emph{whether or not they were lensed images of the same event}.

\paragraph{3. The localizations of the two events rule out astrophysically realistic separations.} Neither localization is very precise, but they are sufficient to constrain the separation of the two unknown sky positions to be $\sim 10\arcdeg$ and rule out sub-arcsecond to sub-arcminute scale separation that would be expected from an astrophysical strong lensing systems.
The localizations are entirely inconsistent with each other, and have no significant overlap.
In simulations of lensed pairs of events with the same separation in time, we find that there is less than one chance in a thousand of obtaining two localizations that are as inconsistent as S190828j and S190828l, due only to measurement uncertainty.
On the other hand, the separation is quite ordinary if we interpret the signals as two unrelated mergers.

\section{Conclusion}

Neither the intrinsic (mass and spin) parameters of these two events, nor the strain data surrounding them, are publicly available yet.
When the waveform parameters or the strain data become available, one can perform additional tests by comparing their waveforms as has been done for previous events by \citet{2019ApJ...874L...2H}.
When the full strain data for all of the present observing run becomes available, one can perform a sub-threshold, mass-constrained search for additional signals with the same waveforms \citep{2019arXiv190406020L}.

It has been argued that a large fraction of LIGO/Virgo \ac{BBH} signals that have been detected to date may be strongly lensed images of much more distance mergers.
Although our results are neutral in that debate, they do show that \emph{S190828j and S190828l were probably not lensed images of the same merger}.

\providecommand{\acrolowercase}[1]{\lowercase{#1}}

\begin{acronym}
\acro{2D}[2D]{two\nobreakdashes-dimensional}
\acro{2+1D}[2+1D]{2+1\nobreakdashes--dimensional}
\acro{2MRS}[2MRS]{2MASS Redshift Survey}
\acro{3D}[3D]{three\nobreakdashes-dimensional}
\acro{2MASS}[2MASS]{Two Micron All Sky Survey}
\acro{AdVirgo}[AdVirgo]{Advanced Virgo}
\acro{AMI}[AMI]{Arcminute Microkelvin Imager}
\acro{AGN}[AGN]{active galactic nucleus}
\acroplural{AGN}[AGN\acrolowercase{s}]{active galactic nuclei}
\acro{aLIGO}[aLIGO]{Advanced \acs{LIGO}}
\acro{ASKAP}[ASKAP]{Australian \acl{SKA} Pathfinder}
\acro{ATCA}[ATCA]{Australia Telescope Compact Array}
\acro{ATLAS}[ATLAS]{Asteroid Terrestrial-impact Last Alert System}
\acro{GW}[GW]{gravitational-wave}
\acro{BAT}[BAT]{Burst Alert Telescope\acroextra{ (instrument on \emph{Swift})}}
\acro{BATSE}[BATSE]{Burst and Transient Source Experiment\acroextra{ (instrument on \acs{CGRO})}}
\acro{BAYESTAR}[BAYESTAR]{BAYESian TriAngulation and Rapid localization}
\acro{BBH}[BBH]{binary black hole}
\acro{BHBH}[BHBH]{\acl{BH}\nobreakdashes--\acl{BH}}
\acro{BH}[BH]{black hole}
\acro{BNS}[BNS]{binary neutron star}
\acro{CARMA}[CARMA]{Combined Array for Research in Millimeter\nobreakdashes-wave Astronomy}
\acro{CASA}[CASA]{Common Astronomy Software Applications}
\acro{CBCG}[CBCG]{Compact Binary Coalescence Galaxy}
\acro{CFH12k}[CFH12k]{Canada--France--Hawaii $12\,288 \times 8\,192$ pixel CCD mosaic\acroextra{ (instrument formerly on the Canada--France--Hawaii Telescope, now on the \ac{P48})}}
\acro{CLU}[CLU]{Census of the Local Universe}
\acro{CRTS}[CRTS]{Catalina Real-time Transient Survey}
\acro{CTIO}[CTIO]{Cerro Tololo Inter-American Observatory}
\acro{CBC}[CBC]{compact binary coalescence}
\acro{CCD}[CCD]{charge coupled device}
\acro{CDF}[CDF]{cumulative distribution function}
\acro{CGRO}[CGRO]{Compton Gamma Ray Observatory}
\acro{CMB}[CMB]{cosmic microwave background}
\acro{CRLB}[CRLB]{Cram\'{e}r\nobreakdashes--Rao lower bound}
\acro{cWB}[\acrolowercase{c}WB]{Coherent WaveBurst}
\acro{DASWG}[DASWG]{Data Analysis Software Working Group}
\acro{DBSP}[DBSP]{Double Spectrograph\acroextra{ (instrument on \acs{P200})}}
\acro{DCT}[DCT]{Discovery Channel Telescope}
\acro{DECAM}[DECam]{Dark Energy Camera\acroextra{ (instrument on the Blanco 4\nobreakdashes-m telescope at \acs{CTIO})}}
\acro{DES}[DES]{Dark Energy Survey}
\acro{DFT}[DFT]{discrete Fourier transform}
\acro{EM}[EM]{electromagnetic}
\acro{ER8}[ER8]{eighth engineering run}
\acro{FD}[FD]{frequency domain}
\acro{FAR}[FAR]{false alarm rate}
\acro{FFT}[FFT]{fast Fourier transform}
\acro{FIR}[FIR]{finite impulse response}
\acro{FITS}[FITS]{Flexible Image Transport System}
\acro{F2}[F2]{FLAMINGOS\nobreakdashes-2}
\acro{FLOPS}[FLOPS]{floating point operations per second}
\acro{FOV}[FOV]{field of view}
\acroplural{FOV}[FOV\acrolowercase{s}]{fields of view}
\acro{FTN}[FTN]{Faulkes Telescope North}
\acro{FWHM}[FWHM]{full width at half-maximum}
\acro{GBM}[GBM]{Gamma-ray Burst Monitor\acroextra{ (instrument on \emph{Fermi})}}
\acro{GCN}[GCN]{Gamma-ray Coordinates Network}
\acro{GLADE}[GLADE]{Galaxy List for the Advanced Detector Era}
\acro{GMOS}[GMOS]{Gemini Multi-object Spectrograph\acroextra{ (instrument on the Gemini telescopes)}}
\acro{GRB}[GRB]{gamma-ray burst}
\acro{GROWTH}[GROWTH]{Global Relay of Observatories Watching Transients Happen}
\acro{GSC}[GSC]{Gas Slit Camera}
\acro{GSL}[GSL]{GNU Scientific Library}
\acro{GTC}[GTC]{Gran Telescopio Canarias}
\acro{GW}[GW]{gravitational wave}
\acro{GWGC}[GWGC]{Gravitational Wave Galaxy Catalogue}
\acro{HAWC}[HAWC]{High\nobreakdashes-Altitude Water \v{C}erenkov Gamma\nobreakdashes-Ray Observatory}
\acro{HCT}[HCT]{Himalayan Chandra Telescope}
\acro{HEALPix}[HEALP\acrolowercase{ix}]{Hierarchical Equal Area isoLatitude Pixelization}
\acro{HEASARC}[HEASARC]{High Energy Astrophysics Science Archive Research Center}
\acro{HETE}[HETE]{High Energy Transient Explorer}
\acro{HFOSC}[HFOSC]{Himalaya Faint Object Spectrograph and Camera\acroextra{ (instrument on \acs{HCT})}}
\acro{HMXB}[HMXB]{high\nobreakdashes-mass X\nobreakdashes-ray binary}
\acroplural{HMXB}[HMXB\acrolowercase{s}]{high\nobreakdashes-mass X\nobreakdashes-ray binaries}
\acro{HSC}[HSC]{Hyper Suprime\nobreakdashes-Cam\acroextra{ (instrument on the 8.2\nobreakdashes-m Subaru telescope)}}
\acro{IACT}[IACT]{imaging atmospheric \v{C}erenkov telescope}
\acro{IIR}[IIR]{infinite impulse response}
\acro{IMACS}[IMACS]{Inamori-Magellan Areal Camera \& Spectrograph\acroextra{ (instrument on the Magellan Baade telescope)}}
\acro{IMR}[IMR]{inspiral-merger-ringdown}
\acro{IPAC}[IPAC]{Infrared Processing and Analysis Center}
\acro{IPN}[IPN]{InterPlanetary Network}
\acro{IPTF}[\acrolowercase{i}PTF]{intermediate \acl{PTF}}
\acro{IRAC}[IRAC]{Infrared Array Camera}
\acro{ISM}[ISM]{interstellar medium}
\acro{ISS}[ISS]{International Space Station}
\acro{KAGRA}[KAGRA]{KAmioka GRAvitational\nobreakdashes-wave observatory}
\acro{KDE}[KDE]{kernel density estimator}
\acro{KN}[KN]{kilonova}
\acroplural{KN}[KNe]{kilonovae}
\acro{LAT}[LAT]{Large Area Telescope}
\acro{LCOGT}[LCOGT]{Las Cumbres Observatory Global Telescope}
\acro{LHO}[LHO]{\ac{LIGO} Hanford Observatory}
\acro{LIB}[LIB]{LALInference Burst}
\acro{LIGO}[LIGO]{Laser Interferometer \acs{GW} Observatory}
\acro{llGRB}[\acrolowercase{ll}GRB]{low\nobreakdashes-luminosity \ac{GRB}}
\acro{LLOID}[LLOID]{Low Latency Online Inspiral Detection}
\acro{LLO}[LLO]{\ac{LIGO} Livingston Observatory}
\acro{LMI}[LMI]{Large Monolithic Imager\acroextra{ (instrument on \ac{DCT})}}
\acro{LOFAR}[LOFAR]{Low Frequency Array}
\acro{LOS}[LOS]{line of sight}
\acroplural{LOS}[LOSs]{lines of sight}
\acro{LMC}[LMC]{Large Magellanic Cloud}
\acro{LSB}[LSB]{long, soft burst}
\acro{LSC}[LSC]{\acs{LIGO} Scientific Collaboration}
\acro{LSO}[LSO]{last stable orbit}
\acro{LSST}[LSST]{Large Synoptic Survey Telescope}
\acro{LT}[LT]{Liverpool Telescope}
\acro{LTI}[LTI]{linear time invariant}
\acro{MAP}[MAP]{maximum a posteriori}
\acro{MBTA}[MBTA]{Multi-Band Template Analysis}
\acro{MCMC}[MCMC]{Markov chain Monte Carlo}
\acro{MLE}[MLE]{\ac{ML} estimator}
\acro{ML}[ML]{maximum likelihood}
\acro{MOU}[MOU]{memorandum of understanding}
\acroplural{MOU}[MOUs]{memoranda of understanding}
\acro{MWA}[MWA]{Murchison Widefield Array}
\acro{NED}[NED]{NASA/IPAC Extragalactic Database}
\acro{NIR}[NIR]{near infrared}
\acro{NSBH}[NSBH]{neutron star\nobreakdashes--black hole}
\acro{NSBH}[NSBH]{\acl{NS}\nobreakdashes--\acl{BH}}
\acro{NSF}[NSF]{National Science Foundation}
\acro{NSNS}[NSNS]{\acl{NS}\nobreakdashes--\acl{NS}}
\acro{NS}[NS]{neutron star}
\acro{O1}[O1]{\acl{aLIGO}'s first observing run}
\acro{O2}[O2]{\acl{aLIGO}'s second observing run}
\acro{O3}[O3]{\acl{aLIGO}'s and \acl{AdVirgo} third observing run}
\acro{oLIB}[\acrolowercase{o}LIB]{Omicron+\acl{LIB}}
\acro{OT}[OT]{optical transient}
\acro{P48}[P48]{Palomar 48~inch Oschin telescope}
\acro{P60}[P60]{robotic Palomar 60~inch telescope}
\acro{P200}[P200]{Palomar 200~inch Hale telescope}
\acro{PC}[PC]{photon counting}
\acro{PESSTO}[PESSTO]{Public ESO Spectroscopic Survey of Transient Objects}
\acro{PSD}[PSD]{power spectral density}
\acro{PSF}[PSF]{point-spread function}
\acro{PS1}[PS1]{Pan\nobreakdashes-STARRS~1}
\acro{PTF}[PTF]{Palomar Transient Factory}
\acro{QUEST}[QUEST]{Quasar Equatorial Survey Team}
\acro{RAPTOR}[RAPTOR]{Rapid Telescopes for Optical Response}
\acro{REU}[REU]{Research Experiences for Undergraduates}
\acro{RMS}[RMS]{root mean square}
\acro{ROTSE}[ROTSE]{Robotic Optical Transient Search}
\acro{S5}[S5]{\ac{LIGO}'s fifth science run}
\acro{S6}[S6]{\ac{LIGO}'s sixth science run}
\acro{SAA}[SAA]{South Atlantic Anomaly}
\acro{SHB}[SHB]{short, hard burst}
\acro{SHGRB}[SHGRB]{short, hard \acl{GRB}}
\acro{SKA}[SKA]{Square Kilometer Array}
\acro{SMT}[SMT]{Slewing Mirror Telescope\acroextra{ (instrument on \acs{UFFO} Pathfinder)}}
\acro{SNR}[S/N]{signal\nobreakdashes-to\nobreakdashes-noise ratio}
\acro{SSC}[SSC]{synchrotron self\nobreakdashes-Compton}
\acro{SDSS}[SDSS]{Sloan Digital Sky Survey}
\acro{SED}[SED]{spectral energy distribution}
\acro{SFR}[SFR]{star formation rate}
\acro{SGRB}[SGRB]{short \acl{GRB}}
\acro{SN}[SN]{supernova}
\acroplural{SN}[SN\acrolowercase{e}]{supernova}
\acro{SNIa}[\acs{SN}\,I\acrolowercase{a}]{Type~Ia \ac{SN}}
\acroplural{SNIa}[\acsp{SN}\,I\acrolowercase{a}]{Type~Ic \acp{SN}}
\acro{SNIcBL}[\acs{SN}\,I\acrolowercase{c}\nobreakdashes-BL]{broad\nobreakdashes-line Type~Ic \ac{SN}}
\acroplural{SNIcBL}[\acsp{SN}\,I\acrolowercase{c}\nobreakdashes-BL]{broad\nobreakdashes-line Type~Ic \acp{SN}}
\acro{SVD}[SVD]{singular value decomposition}
\acro{TAROT}[TAROT]{T\'{e}lescopes \`{a} Action Rapide pour les Objets Transitoires}
\acro{TDOA}[TDOA]{time delay on arrival}
\acroplural{TDOA}[TDOA\acrolowercase{s}]{time delays on arrival}
\acro{TD}[TD]{time domain}
\acro{TOA}[TOA]{time of arrival}
\acroplural{TOA}[TOA\acrolowercase{s}]{times of arrival}
\acro{TOO}[TOO]{target\nobreakdashes-of\nobreakdashes-opportunity}
\acroplural{TOO}[TOO\acrolowercase{s}]{targets of opportunity}
\acro{UFFO}[UFFO]{Ultra Fast Flash Observatory}
\acro{UHE}[UHE]{ultra high energy}
\acro{UVOT}[UVOT]{UV/Optical Telescope\acroextra{ (instrument on \emph{Swift})}}
\acro{VHE}[VHE]{very high energy}
\acro{VISTA}[VISTA@ESO]{Visible and Infrared Survey Telescope}
\acro{VLA}[VLA]{Karl G. Jansky Very Large Array}
\acro{VLT}[VLT]{Very Large Telescope}
\acro{VST}[VST@ESO]{\acs{VLT} Survey Telescope}
\acro{WAM}[WAM]{Wide\nobreakdashes-band All\nobreakdashes-sky Monitor\acroextra{ (instrument on \emph{Suzaku})}}
\acro{WCS}[WCS]{World Coordinate System}
\acro{WSS}[w.s.s.]{wide\nobreakdashes-sense stationary}
\acro{XRF}[XRF]{X\nobreakdashes-ray flash}
\acroplural{XRF}[XRF\acrolowercase{s}]{X\nobreakdashes-ray flashes}
\acro{XRT}[XRT]{X\nobreakdashes-ray Telescope\acroextra{ (instrument on \emph{Swift})}}
\acro{ZTF}[ZTF]{Zwicky Transient Facility}
\end{acronym}

\acknowledgments

We thank Ariel Goobar, Albert Kong, and Daniel Mortlock for fruitful discussions.

Much of this work was performed at the Aspen Center for Physics, which is supported by National Science Foundation grant PHY-1607611.

This work was supported by the GROWTH (Global Relay of Observatories Watching Transients Happen) project funded by the National Science Foundation under PIRE Grant No. 1545949.

DAG acknowledges support from Hubble Fellowship grant HST-HF2-51408.001-A.
Support for Program number HST-HF2-51408.001-A is provided by NASA
through a grant from the Space Telescope Science Institute, which is operated by the Association of Universities for Research in Astronomy, Incorporated, under NASA contract NAS5-26555.

JSB was partially supported by a Gordon and Betty Moore Foundation Data-Driven Discovery grant.

We gratefully acknowledge Amazon Web Services, Inc. for a generous grant (\texttt{PS\_IK\_FY2019\_Q3\_ Caltech\_Gravitational\_Wave}) that funded our use of the Amazon Web Services cloud computing infrastructure to perform the simulations.

\facilities{
    LIGO,
    EGO:Virgo
}

\software{
    astropy \citep{2013A&A...558A..33A},
    HEALPix \citep{2005ApJ...622..759G},
    Healpy \citep{2019JOSS....4.1298Z},
    ligo.skymap (\url{https://git.ligo.org/lscsoft/ligo.skymap})
}

\bibliography{main}{}
\bibliographystyle{aasjournal}

\end{document}